\documentclass{emulateapj}   
   
\shorttitle{Peak energy clustering} \shortauthors{Peer, M\'esz\'aros \& Rees}

\newcommand{\eV}{\rm{\, eV }}   
\newcommand{\keV}{\rm{\, keV }}   
\newcommand{\MeV}{\rm{\, MeV }}

\newcommand{\beq}{\begin{equation}}   
\newcommand{\eeq}{\end{equation}}   
\newcommand{\ba}{\begin{array}}   
\newcommand{\ea}{\end{array}}   
  
\newcommand{\Gi}{\Gamma_{2}}  
  
\newcommand{\ee}{\epsilon_{e,-0.5}}  
  
\newcommand{\ed}{\epsilon_{d,-1}}  
  
\def \etal{{\it et al.~}}  
  
\begin{document}   
   
\title{Peak energy clustering and efficiency in compact objects}   
   
\author{Asaf Pe'er\altaffilmark{1}\altaffilmark{2}\altaffilmark{3}, Peter M\'esz\'aros\altaffilmark{2} and Martin J. Rees\altaffilmark{4} }   
\altaffiltext{1}{Astronomical Institute ``Anton Pannekoek'', Kruislaan 403, 1098SJ Amsterdam, the Netherlands}   
\altaffiltext{2}{Dpt. Astron. \& Astrophysics, Dpt. Physics, Pennsylvania State University, University Park, PA 16802}  
\altaffiltext{3}{apeer@science.uva.nl}  
\altaffiltext{4}{Institute of Astronomy, University of Cambridge, Madingley Rd., Cambridge CB3 0HA, UK}  
  
\begin{abstract}   
We study the properties of plasmas containing a low energy thermal  
photon component at comoving temperature $\theta \equiv kT'/m_e  
c^2 \sim 10^{-5} - 10^{-2}$ interacting with an energetic electron   
component, characteristic of, e.g., the dissipation phase of  
relativistic outflows in gamma-ray bursts (GRB's), X-ray flashes,  
and blazars. We show that, for scattering optical depths larger than   
a few, balance between Compton and inverse-Compton scattering leads to  
the accumulation of electrons at values of $\gamma\beta \sim 0.1 -  
0.3$. For optical depths larger than $\sim 100$, this leads to a  
peak in the comoving photon spectrum at $1-10 \keV$,   
very weakly dependent on the values of the free parameters.  
In particular, these results are applicable to the internal   
shock model of GRB, as well as to slow dissipation  models,  
e.g. as might be expected from reconnection, if the dissipation   
occurs at a sub-photospheric radii.   
For GRB bulk Lorentz factors $\sim 100$, this results in observed   
spectral peaks clustering in the $0.1-1$ MeV range, with conversion   
efficiencies of electron into photon energy in the BATSE range of   
$\sim 30\%$.   
  
\end{abstract}   
   
   
\section{Introduction}  
\label{sec:intro}  
  
The widely accepted interpretation of gamma-ray burst (GRB)  
phenomenology is that the observable radiation is due to the  
dissipation of the kinetic energy of a relativistic outflow,   
powered by a central compact object \citep[for reviews, see,  
e.g.,][]{Mes02,Waxman03,Piran04}.   
The dissipated energy is assumed to be converted to   
energetic electrons, which produce high energy photons by    
synchrotron radiation and inverse Compton (IC) scattering.     
Similar considerations are assumed to apply to X-ray flashes  
(XRFs) and some models of blazars \citep[e.g.,][]{Guetta04}.  
  
Even though the optically thin synchrotron-IC emission model   
of GRB's \citep{Band93,Tavani96,Preece98} and blazars is in   
general agreement with observations, two concerns   
are often raised in particular for GRBs.  
The $\gamma$~-ray break energy of most GRB's observed by BATSE  
is in the range 100~\keV - 300~\keV \citep{Brainerd98, Preece00}. It is   
thought that clustering of the peak emission in this  
narrow energy range requires fine tuning of the fireball model  
parameters. In addition, there is evidence in some bursts   
for low-energy spectral slopes steeper than the optically thin   
synchrotron predictions \citep{Crider97, Preece98, Frontera00,  
  Ghirlanda03}.   
  
This has led several authors to consider alternative scenarios   
for the $\gamma$-ray emission, such as Compton drag \citep{LGCR00},   
or the contribution of a photospheric term \citep{MR00,MRRZ,MR04}  
{\citep[but see also][]{SP04}}.   
In the {``photospheric''} model, thermal radiation, originating  
near the base of    
the flow emerges from a ``photospheric radius" at which the flow becomes  
optically thin. As shown in \citet{MR04}, dissipation  
processes can occur at radii smaller than the photosphere radius,  
increasing the energy available to produce thermal seed photons   
which can be IC scattered by relativistic electrons. The high  
optical depth to scattering by the electrons (and the created pairs)  
results in a spectrum that is significantly different than the  
optically thin synchrotron-SSC model predictions.   
  
The spectrum from internal collisions at a high scattering optical   
depth was studied by \citet{PW04}, with a numerical model which calculates   
self-consistently the photon and electron distributions \citep{PW05}.   
Although a thermal component was not considered in this work, it  
was found that a large number of pairs can be created, and for  
optical depths $\tau_\pm\sim 100$, the emission peaks at $\sim 1 \MeV$.   
This result was found to be very robust, with a very weak dependence on   
the values of the free parameters.   
  
In this letter, we analyze numerically and analytically the emission   
resulting from the interaction of energetic electrons with a low   
energy thermal photon component, under conditions of intermediate to   
large scattering optical depths, $\tau \approx 10 - 100$.    
We show in \S\ref{sec:peak_energy} that the relativistic electrons  
lose their energy and accumulate around (comoving) momentum values   
$\gamma \beta \approx 0.1 - 0.3$. This results from a balance   
between inverse-Compton cooling and Compton heating, which is  
very insensitive to other input parameters. For a bulk Lorentz factor  
$\Gamma\sim 10^2$ as observed in GRB (or $\Gamma\sim 10-50$ in blazars)   
an observed Wien peak at $\lesssim 1 \MeV$ is thus obtained.  
In \S\ref{sec:model} we specify a model for the GRB prompt emission,  
in which part of the dissipation or internal collisions occur near or  
below the photospheric radius.  
We show in \S\ref{sec:efficiency} that using the commonly assumed  
values of the free parameters of the fireball model, a spectral  peak   
at $\lesssim \MeV$ is obtained, with a radiative efficiency which can   
be substantially higher than for standard internal shocks.   
We also consider a ``slow dissipation" case, which might characterize   
magnetic dissipation models, where electrons are reenergized to lower   
energies multiple times, leading to a similar peak energy clustering   
and efficiency.  We discuss in \S\ref{sec:discussion} the implications   
for GRB, XRF and AGN.

\section{Electron energy balance }  
\label{sec:peak_energy}  
  
We consider a plasma cloud of characteristic comoving length $\Delta$,   
containing a thermal component of low energy photons at a normalized   
comoving  
temperature $\theta \equiv kT'/m_ec^2$, with an external source (e.g., shocks   
or other dissipative processes) which continuously injects into the plasma  
isotropically distributed electrons at a characteristic Lorentz factor   
$\gamma_m$, at a constant rate during the comoving dynamical time   
$t_{dyn} \equiv \Delta / c$. The initial energy density of the photon component,   
$u_{ph}$, is parametrized relative to the injected electron energy density   
$u_{el}$ through the numerical factor $A=u_{ph}/u_{el}$. For models of GRBs  
and XRFs, if the dissipative process occurs near or below the Thompson   
photosphere, a typical value is $A\sim 1$ (see  
equ. \ref{eq:epsilon_f}).   
{ For $A> \theta / \gamma_m \sim 10^{-4}$, the occupation number of  
  up-scattered photons is much smaller than unity, and induced  
  scattering can be ignored.}   
We assume (I) that $\gamma_m \theta < 1$, i.e. that the scattering is   
in the Thompson regime; and (II) that $\gamma_m^2 \theta > 1$, i.e., that  
electrons scatter photons to energies above the electron's rest mass energy.  
  
If the main electron energy loss is inverse-Compton scattering, the   
injected electrons lose energy at an initial rate $ \tilde{t}_{loss} =   
{\gamma_m m_e c^2}/{(4/3) c \sigma_T \gamma_m^2 u_{ph}} $,   
and cool  down to $\gamma_f \simeq 1$ on a total loss time     
$t_{loss} = \tilde{t}_{loss}\times \gamma_m$, therefore  
\beq  
{t_{loss} \over t_{dyn}} = {3 \over 4 c \sigma_T A n_{el} \gamma_m   
t_{dyn}} = {3 \over 4 A \gamma_m \tau_{\gamma e}}.  
\label{eq:t_loss}  
\eeq  
Here $n_{el} = u_{el}/\gamma_m m_e c^2$, and $\tau_{\gamma e}$ is the  
electron scattering optical depth.  
Thus, for $\tau_{\gamma e} > 1$,   
electrons accumulate at $\gamma_f \approx 1$ on a time shorter than  
 the dynamical time.   
  
For concreteness we focus below on the internal shock case   
\citep[the qualitatively similar slow dissipation case is discussed in   
more detail in][] {PM05}.  
The cooled electrons form a steady state power law distribution   
in the range $\gamma_f < \gamma < \gamma_m$, of the form { 
$n_{el}(\gamma\beta) \propto (\gamma\beta)^{-2}$, where $\beta =  
(1-\gamma^{-2})^{1/2}$}.  The emitted photon spectrum   
in the range $3\theta m_e c^2 < \varepsilon <  4 \theta \gamma_m^2   
m_e c^2$ is $n_\gamma(\varepsilon) \propto \varepsilon^{-3/2}$,   
thus the characteristic energy of photons more energetic than $m_e  
c^2$ is not far above $m_e c^2$. These photons receive nearly all of   
the energy of the injected electrons.

The value of the electron momenta at the end of the dynamical time, 
$\gamma_f \beta_f$, can now be calculated as a function of the optical depth,  
the initial photon temperature $\theta$, and the ratio of energies $A$.    
This value is estimated by balance between energy loss due to  
inverse-Compton scattering of soft photons at energies   
$\varepsilon < m_e c^2$, and energy gain by direct Compton scattering  
photons more energetic than $m_e c^2$.   
As we show below, the weak dependence of $\gamma_f \beta_f$ on any of  
the values of the free parameters is a key result that implies weak  
dependence of the observed spectral peak on the values of the free  
parameters, for intermediate to high values of the optical depth.

A photon with energy $\varepsilon > m_e c^2$, undergoing  
Compton scattering with an isotropically distributed electron  
density $n_{el}$ with sub-relativistic velocity $\gamma \simeq 1$,    
loses energy at a constant   
rate\footnote{See http://staff.science.uva.nl/$\sim$apeer; A  
  derivative of this result also appears in \citet{PM05}.}  
\footnote{Roughly, Klein-Nishina Compton cross section is  
$\sigma(\varepsilon) \approx (3/16) \sigma_T \times m_e c^2  
\varepsilon^{-1}$, thus  
$d\varepsilon/dt \approx c \sigma_T n_{el} m_e c^2 \varepsilon^0$. 
  Full analytical treatment gives the numerical pre-factor 1/2.}  
\beq  
{d \varepsilon \over dt} \simeq - c \sigma_T (m_e c^2 / 2) n_{el}.  
\eeq  
This { result} is valid for photon energies $\gtrsim f m_e c^2$, where    
$f \simeq 3$. 
The injection rate by IC scattering of photons   
at energies $\gtrsim f m_e c^2$  is approximated by   
$dn_{ph}(\varepsilon > f m_e c^2)/dt \approx u_{el} / f m_e c^2 t_{dyn}$.  
These photons lose energy by down-scattering to energies  
below $f m_e c^2$ on a time scale   
$\varepsilon/(d\varepsilon/dt) = 2 f t_{dyn} / \tau_{\gamma e}$, which  
is shorter than the dynamical time for $\tau_{\gamma e} > 2 f \simeq  
6$. Assuming $\tau_{\gamma e} > few$, the photon's number density  
approaches a steady value given by
\beq  
n_{ph}(\varepsilon \gtrsim f m_e c^2) \approx { u_{el} \over m_e c^2} {2  
\over \tau_{\gamma e}}.   
\label{eq:n_ph_up}  
\eeq  
The rate of energy gain by an electron at $\gamma_f$ due to  
Compton scattering of these photons is { therefore}  
\beq  
{d E_+ \over dt} \approx {m_e c^2 \over 2} c \sigma_T {2 u_{el} \over  
m_e c^2 \tau_{\gamma e}}.  
\label{eq:dE+}  
\eeq

In the Thompson regime, which is valid for photons at energies  
$\varepsilon(n_{sc}) \lesssim (\gamma_f \beta_f)^2 m_e c^2 / f$ with $f  
\approx 3$, electrons at $\gamma_f$ up-scatter the low energy thermal  
photons at a rate $\approx n_{ph} c \sigma_T$, where $n_{ph} \approx  
u_{ph}/3 \theta m_e c^2$.    
In this regime, the increase in the photon energy at each scattering is  
$\Delta \varepsilon = (4/3) (\gamma_f \beta_f)^2 \varepsilon(n_{sc})$,  
and the photon energy after $n_{sc}$ scattering is   
$\varepsilon(n_{sc}) \approx 3 \theta m_e c^2 \exp(4/3 (\gamma_f  
\beta_f)^2 n_{sc})$ where $\beta_f \equiv (1 - \gamma_f^{-2})^{1/2}$.   
For $\tau_{\gamma e} > few$, nearly all of the photons are being  
scattered and do not leave the plasma during the dynamical time,  
therefore the electron energy loss rate is time independent, given by          
\beq  
{d E_- \over dt} \approx (4/3) (\gamma_f \beta_f)^2  c \sigma_T u_{ph}  
e^{4/3 (\gamma_f \beta_f)^2 n_{sc}}.   
\eeq  
  
Equating the energy loss and energy gain rates,   
\beq  
(\gamma_f \beta_f)^2 e^{4/3 (\gamma_f \beta_f)^2 n_{sc}} = {3 \over 4  
A \tau_{\gamma e}}.  
\label{eq:gb1}  
\eeq  
  
For optical depths not much larger than a few, the exponent on the   
right hand side of equation \ref{eq:gb1} can be approximated as 1,  
since for a relativistically expanding plasma $\tau_{\gamma e} \simeq  
n_{sc}$, and $\gamma_f \beta_f < 1$. In this approximation, the steady  
state electron momentum is   
given by    
\beq  
\gamma_f \beta_f (n_{sc} \lesssim 10) \approx \left({ 3 \over 4 A   
\tau_{\gamma e}}\right)^{1/2} \approx \, 0.3 \,  
A_0^{-1/2} \tau_{\gamma e,1}^{-1/2},  
\label{eq:gb_f_low}  
\eeq  
where $\tau_{\gamma e}=10^1 \tau_{\gamma e,1}$ assumed.  
The spectral peak after an intermediate number of scatterings is thus  
expected between $\theta m_e c^2$ and $(\gamma_f \beta_f)^2 m_e c^2$.  
  
For optical depths larger than a few tens, low energy photons can be  
upscattered to a maximum energy $(\gamma_f \beta_f)^2 m_e c^2 / f $  
in the Thompson regime, and the exponent in equation \ref{eq:gb1} is   
approximated by $\exp(4/3 (\gamma_f \beta_f)^2 n_{sc}) \approx  
(\gamma_f \beta_f)^2/ 3\theta f$. The steady state electron momentum  
is therefore     
\beq  
\gamma_f \beta_f \approx \left( {9 \theta f \over 4 A \tau_{\gamma e}}  
\right)^{1/4} \approx \, 0.1 \, \theta_{-3}^{1/4} A_0^{-1/4}  
\tau_{\gamma e,2}^{-1/4},  
\label{eq:gb_f}  
\eeq  
where $\theta =10^{-3}\theta_{-3}$, $A=10^0 A_0$,  and  
$\tau_{\gamma e}=10^2 \tau_{\gamma e,2}$ were assumed.   
The number of scatterings required for photons to be upscattered to  
$(\gamma_f \beta_f)^2 m_e c^2 / f$ can be estimated as    
\beq  
n_{sc} \approx \frac{\log\left({ (\gamma_f \beta_f)^2 \over  
3 \theta f}\right)}{{4 \over 3} (\gamma_f \beta_f)^2} \approx 10^{1.5} -  
10^{2.5},  
\label{eq:n_sc}  
\eeq   
for $\theta$ in the range $10^{-5} - 10^{-2}$. For this number of  
scatterings, the spectral peak is expected at $(\gamma_f \beta_f)^2 m_e  
c^2 /f \simeq 1-3 \keV$ (in the plasma frame). Both this result and  
the result in eq. \ref{eq:gb_f} show a very weak dependence on any of  
the parameter values.  
  
For optical depths higher than the values given in eq. \ref{eq:n_sc},  
photons are upscattered to energies above $(\gamma_f \beta_f)^2 m_e  
c^2 / f$ outside the Thompson regime. The final photon energy is  
therefore limited to a narrow range,$(\gamma_f \beta_f)^2 m_e   
c^2 / f \leq \varepsilon(n_{sc}) \leq (\gamma_f \beta_f)^2 m_ec^2$,  
and a Wien peak is formed. Since the average photon energy   
is equal to the average kinetic energy of the electron in this case,  
a Compton equilibrium is formed \citep{GFR83, LZ87, S87},   
the final electron momentum is   
\beq  
\gamma_f \beta_f = \left[ 3 \theta \left(1 +  
  A^{-1}\right)\right]^{1/2} = 0.08 \, \theta_{-3}^{1/2},  
\eeq  
and a Wien peak is formed at   
\beq  
\varepsilon_{WP} \approx 3 \theta m_e c^2 \times \left(1 + A^{-1}  
\right) \simeq 10 \keV,   
\eeq  
irrespective of the value of the optical depth.

Pairs produced by $\gamma\gamma$ will accumulate at   
$\gamma_f \beta_f$, while lowering of the number density of   
energetic photons, hence lowering the value of $\gamma_f \beta_f$.   
We estimate the significance of this effect using the pair production  
rate of photons at energy $\varepsilon$, $dt(\varepsilon)^{-1}  
\approx (1/4) c \sigma_T n_{ph}(\tilde\varepsilon \gtrsim  
(m_ec^2)^2/\varepsilon)$.   
The ratio of the characteristic times for pair production and Compton  
scattering by electrons at $\gamma_f \simeq 1$, for   
photons at energy $f m_e c^2$ is   
\beq  
{\left.t_{loss}\right|_{p.p} \over \left.t_{loss}\right|_{C}} \simeq {n_{el}  
\over f n_{ph}\left(\tilde\epsilon\gtrsim {m_e c^2 \over f}\right)}.   
\eeq  
If Compton scattering is the only mechanism producing photons   
above the thermal peak, then   
$n_{ph}(\tilde\epsilon\gtrsim m_e c^2/f) \approx n_{el} \times (f/\theta)^{1/2}$.  
In this case, the number density of photons at $\gtrsim f m_e c^2$  
is given by equation \ref{eq:n_ph_up}, corrected by a factor  
$2 \theta^{1/2} f^{-3/2} = 3.8 \times 10^{-2} \theta_{-2}^{1/2}  
f_{0.5}^{-3/2}$, where $f=10^{0.5} f_{0.5}$.  
This correction factor to the $1/4$ power enters equation  
\ref{eq:gb_f}, modifying the value of $\gamma_f \beta_f$ in equation  
\ref{eq:gb_f} by a factor $\sim 2$.   
Pair annihilation significantly lowers the effect of pairs on the  
value of $\gamma_f \beta_f$  \citep[see][]{PW04}.  
  
Synchrotron emission affects several aspects of the calculation.   
First, the electron cooling time is shorter than the estimate  
of equation \ref{eq:t_loss}, which neglected synchrotron radiation.   
Second, the photon spectrum is different than the pure Compton   
spectrum calculated above, and depends on the details of the   
electron injection spectrum. Adopting the commonly used power law energy   
distribution of electrons injected above $\gamma_m$, one obtains a   
power law spectrum of synchrotron photons between photon  
energies $\varepsilon_{syn,min}$ and $\varepsilon_{syn,max}$.   
GRB observations suggest that if the low energy component is due to   
synchrotron emission, then  $\varepsilon_{syn.min} \simeq   
10^{-5} - 10^{-1} \times m_e c^2$, i.e., of the same order as   
the assumed value of $\theta$  \citep{Tavani96, Frontera00}.  
Assuming $A=u_{ph}/u_{el} \sim 1$, the photon density around $m_ec^2$   
differs from the estimates above by factors $\lesssim$ few, which   
enters equation \ref{eq:gb_f} to the power $1/4$.   
Finally, the synchrotron self-absorption would increase   
$\gamma_f \beta_f$, by an amount depending on spectral details.    
Numerical calculations for various models \citep{GGS88, PW04,PM05}   
suggest that it does not change the $\gamma_f \beta_f$ of equation   
\ref{eq:gb_f} by more than a factor of a few.    
  
\section{Application to sub-photospheric dissipation}  
\label{sec:model}  
  
As an application, we consider the effect of a photospheric   
term in the gamma-ray burst prompt emission \citep[see][]{MR00, MR04}.    
Following \citet{MR04} we assume that the dissipation occurs at   
$r>r_s \equiv \eta r_0$, where the observed photospheric luminosity is   
$L_\gamma (r)= L_0(r/r_s)^{-2/3}$. Here, $L_0$ is the total luminosity,   
$r_0 = \alpha r_g$ is the size at the base of the flow,   
$r_s$ is the saturation radius where the bulk Lorentz factor   
asymptotes to $\eta$, $r_g$ is the Schwarzschild radius of the   
central object, and $\alpha \geq 1$.   
For an internal shock model of GRB's, variations of the flow   
$\Delta \Gamma \sim \Gamma$ on a minimum time scale $\Delta t \sim r_0/c$   
result in the development of shocks at a minimum radius $r_i \approx  
2 \Gamma r_s$, where $\Gamma \approx \eta$, thus  $L_\gamma  
(r_i)=L_0(2\Gamma)^{-2/3}$.   
The comoving proton density is  
$n_p\approx L_0/4\pi r_i^2 c \Gamma^2 m_p c^2$, and the Thompson  
optical depth due to baryon related electrons at $r_i$ is  
  
\beq  
\tau_{\gamma e} = \Gamma r_0 n_p \sigma_T = 100 \, L_{52}  
\Gamma_2^{-5} (\alpha m_1)^{-1},  
\label{eq:tau}  
\eeq  
where $L_0 = 10^{52}\, L_{52} {\rm\, erg\,s^{-1}}$, $\Gamma = 10^2  
\Gamma_2$, and $M\sim 10 m_1$ solar masses for the central object    
(e.g., black hole). Thus, for the parameters characterizing GRB's,  
the minimum shock dissipation radius can indeed be $r_i < r_{ph}$.  
  
The normalized comoving temperature at $r_i$ is   
\beq  
\ba{ll}  
\theta & \equiv {k_B T' \over m_e c^2} = {k_B \over m_e c^2}\left( \frac{L_0}{4 \pi r_s^2 \Gamma^2 c a}  
  \right)^{1/4} ( 2 \Gamma)^{-2/3} \nonumber \\ & =1.2 \times 10^{-3} L_{52}^{1/4} \Gamma_2^{-5/3}  
  (\alpha m_1)^{-1/2},   
\label{eq:temp}  
\ea  
\eeq  
where $k_B$, $a$ are Bolzmann's and Stefan's constants.  
The shock waves (or generic dissipation mechanism) dissipate some   
fraction $\epsilon_d<1$ of the kinetic energy, $L_k \sim L_0$, resulting   
in an internal (dissipated) energy density   
$u_{int} = {L_0 \epsilon_d}/{4 \pi r_i^2 c \Gamma^2}$.    
The electrons receive a fraction $\epsilon_e$ of this energy,   
so the ratio of the thermal photon energy density $u_{ph}=aT'^4$   
to the electron energy density is   
\beq  
A \equiv {u_{ph} \over u_{el}} =  0.44 \Gi^{-2/3} \ed^{-1} \ee^{-1},  
\label{eq:epsilon_f}  
\eeq  
where $\epsilon_d = 10^{-1} \ed$ and $\epsilon_e = 10^{-0.5} \ee$.  
  
\section{Spectrum and efficiency}  
\label{sec:efficiency}  
  
We calculated numerically the photon and particle energy  
distribution under the assumptions of \S\ref{sec:model} using   
the time dependent numerical model described in \citet{PW05}, including  
synchrotron emission and absorption, and with the addition of a thermal   
photospheric component \citep[see][]{PM05}.  
  
\begin{figure}  
\plotone{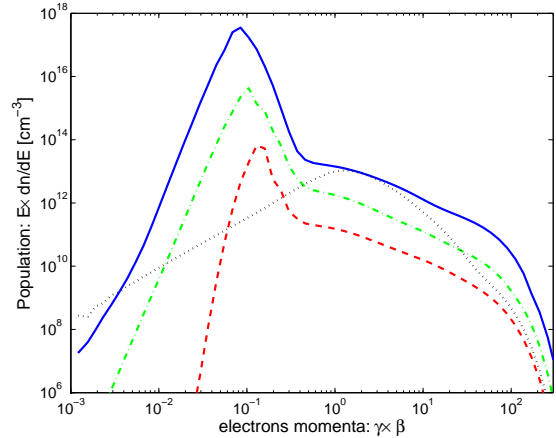}  
\caption{Electron distribution at the end of the dissipation. Results  
are for $L = 10^{52} {\, \rm erg \, s^{-1}}$, $\epsilon_e =  
\epsilon_B = 10^{-0.5}$, $\Gamma = 100$, $\epsilon_d = 0.1$, and  
internals shocks with $\alpha = 1$ (solid line), 10 (dot-dash),   
100 (dash), and Thompson optical depths $\tau_{\gamma e} =   
100, 10, 1$, respectively. The dotted line shows the electrons  
  momenta in the absence of Compton scattering, for $\alpha=1$.}  
\label{fig:elec}  
\end{figure}

The self-consistent electron energy distribution is shown in  
figure~\ref{fig:elec} for three values of the dimensionless lower radius,   
$\alpha = 1,10,100$, i.e. scattering optical depths at the  
minimum dissipation radius of $\tau_{\gamma e} = 100, 10, 1$, respectively   
(see eq. \ref{eq:tau}).  The electrons accumulate at momentum values  
$\gamma_f \beta_f = 0.08, 0.10, 0.14$, in good agreement with  
the analytical calculations of \S\ref{sec:peak_energy}. The  
small deviation (by $\lesssim 2$) between the  
values of equations \ref{eq:gb_f_low}, \ref{eq:gb_f},  and the  
numerical results originates from the synchrotron   
emission ($\epsilon_B = 10^{-0.5}$ in this graph).  
For completeness, we have added a curve showing the electron
distribution in the absence of Compton scattering (i.e., with
synchrotron emission and self absorption only). Synchrotron self
absorption prevents the electrons from cooling below $\gamma \beta
\approx 1$ \citep[see also][]{GGS88}.

\begin{figure}  
\plotone{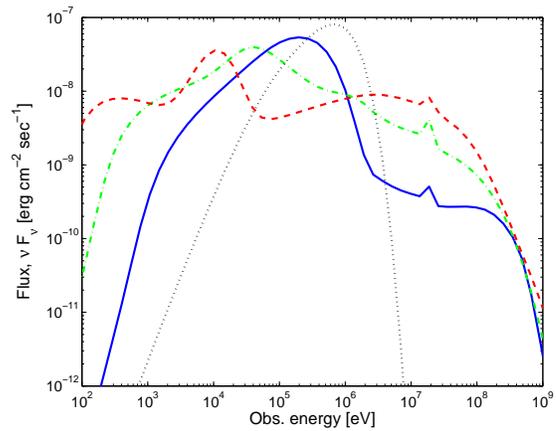}  
\caption{Time averaged spectra for the internal shock cases   
of figure \ref{fig:elec} (solid, dash-dotted and dash-dash),  
and for the slow heating scenario (dots) with similar parameters   
as the solid curve.  A redshift $z=1$ in a flat universe with   
$\Omega_m = 0.3, \Omega_\Lambda = 0.7$, $H_0 = 70$ is assumed.   
The spectrum is not corrected for the energy loss due to adiabatic   
expansion following the dissipation, which lowers the energies by a   
factor 2-3.  
}  
\label{fig:spectrum}  
\end{figure}

The observer-frame photon spectra for the three values of $\alpha$   
(or optical depth) are shown in figure~\ref{fig:spectrum},  
for an assumed bulk Lorentz factor $\Gamma=10^2$. At low optical depths,   
$\tau_{\gamma e} = 1$, the thermal peak at $2.8 \Gamma  
  \theta/(1+z) \approx 10\keV$ is prominent above the    
synchrotron emission component, which dominates the spectrum at  
  low energies, $100 \eV - 10 \keV$. Compton scattering  
  produces the nearly flat spectrum at higher energies, $100 \keV -  
  100 \MeV$.\footnote{Note that since the electrons distribution is  
 affected by numerous physical processes, spectral curves which  
 include both synchrotron and Compton cannot be fully decomposed into  
 synchrotron and Compton components.}  
 The small peaks at $\Gamma m_e c^2/(1+z) \approx 25 \MeV$ are due to
 pair annihilation.   
At $\tau_{\gamma e} = 100$, Comptonization by electrons at $\gamma_f  
\beta_f$ produces the Wien peak at $\varepsilon^{ob.} =   
\Gamma (\gamma_f \beta_f)^2 m_e c^2 / (1+z) = 200 \keV$,   
in accordance with the estimates of \S\ref{sec:peak_energy}.   
Since most of the photons undergo multiple scattering, the  
 Compton component at higher energies is significantly reduced.  
For an intermediate $\tau_{\gamma e} = 10$, the peak is   
higher than $\Gamma \theta m_e c^2$ but lower than   
$\Gamma (\gamma_f \beta_f)^2 m_e c^2$, as expected.   
  
In addition to the sub-photospheric internal shock cases, we   
have also calculated the spectrum in a ``slow heating" scenario,   
for a high optical depth $\tau_{\gamma e}=100$ ($\alpha = 1$).   
In this scenario, the dissipated energy is continuously and equally   
distributed among the electrons in the dissipation region   
\citep{GC99, PW04}. Unlike in previous calculations, here the electrons   
are assumed to be energized under a photosphere, so they interact with   
a strong thermal photon bath. Even though the details of the energy   
injection in this slow scenario are different from those of the internal   
shock scenario, the similar Compton energy balance considerations at  
the high optical depth $\tau_{\gamma e} =100$ result in a Wien peak at   
$\sim 700 \keV$, similar to that of the internal shock scenario. This   
confirms the robustness of this result \citep[see][for details]{PM05}.    
  
For a high optical depth in the dissipation region,  
the photons are trapped with the electrons in the plasma,  
before escaping at $\tau_{\gamma e} \sim 1$. For an adiabatic expansion   
of the plasma following the internal shocks or dissipation occurring  
at $\tau_{\gamma e} = 10 - 100$, \citet{PW04} showed   
that the photons lose $50\% - 70\%$ of their energy before escaping.   
This energy is converted into bulk motion of the expanding plasma.   
We thus expect the observed energy of the Compton peak, produced   
at sub-\MeV, to be lowered by a factor of 2-3 from the values in   
figure~\ref{fig:spectrum}, leading to an observed peak at   
$\approx 100-200 \keV$. This result depends linearly on the unknown   
value of the Lorentz factor $\Gamma$ above the saturation radius.    
   
The accumulation of electrons at $\gamma_f \simeq 1$ implies that  
most of the electron energy injected into the plasma is transfered to   
photons during the dynamical time. For $\tau_{\gamma e} \gtrsim 100$   
at injection, most of this photon energy is near the peak, at an   
observed energy $\Gamma (\gamma_f \beta_f)^2 m_e c^2$.  Since after   
adiabatic expansion the photons maintain $\sim 30\%$ of their energy,   
we conclude that in our model approximately $30\%$ of the energy   
dissipated into electrons is ultimately converted into photons in   
the BATSE range.    
  
\section{Summary and Discussion}  
\label{sec:discussion}  
  
We have considered a plasma composed of a low energy photon component  
and energetic particles, and showed that for Thompson optical depths  
$\tau_{\gamma e} \gtrsim 1$ the electrons accumulate at $\gamma_f \beta_f   
\sim 0.1 - 0.3$, with only a weak dependence on the unknown parameter  
values. For $\tau_{\gamma e} \simeq 100$, the balance between Compton   
and inverse-Compton scattering by these electrons results in a spectral  
peak at $\sim 1-10 \keV$ in the plasma frame. We have presented a   
specific model for GRB prompt emission (\S\ref{sec:model}) in which the   
dissipation occurs below the photosphere, showing the applicability of   
this general result to GRB's.   
  
For both an internal shock model or for a slow dissipation model in   
the presence of a photospheric component, we conclude that if a   
significant part of the energy is dissipated below the photosphere,   
the observed spectral peak clusters naturally in the BATSE range of   
hundreds of keV, for the inferred bulk Lorentz factors $\Gamma\sim 100$.   
This effect is further enhanced by pair creation, e.g. as considered   
in previous analyses \citep{PW04,MR04}. Pairs will accumulate at the   
same $\gamma_f \beta_f$ as the electrons, forming a second photosphere   
at larger radius than the original baryon photosphere. Hence, if the   
dissipation process occurs below the second photosphere, similar results   
are obtained.  
  
The results presented in \S\ref{sec:model} are largely independent   
of the specific model details. As shown in \S\ref{sec:efficiency},  
numerical calculations including magnetic fields lead to results   
close to the analytical predictions. Similar results are obtained   
in  \S\ref{sec:efficiency} for both internal shocks or for a slow   
dissipation model, such as might be expected from magnetic dissipation.   
A similar spectral peak was also found in \citet{PW04}, for an internal  
shock model with synchrotron and SSC emission but no thermal component.   
The dominant effect of a thermal component, as considered here for   
both internal shocks and slow dissipation, is to greatly   
increase the radiative efficiency, to values $\sim 30\%$.  
We thus conclude that a Compton-inverse Compton balance leads to   
the creation of spectral peak at comoving photon energies $1-10$ keV,   
which for inferred bulk Lorentz factors of order $\Gamma \sim   
100\Gamma_2$ leads to observed peaks at $100-1000 \Gamma_2 \keV$,   
with a high efficiency. This is regardless of the nature of the   
dissipation process (shocks, magnetic reconnection, etc),   
provided it occurs at large optical depth, such as near or below   
the Thompson photosphere. If the spectral peak is a Wien peak,   
its observed energy is indicative of the asymptotic value of   
the bulk Lorentz factor $\Gamma$.    
  
These results may be applicable to a range of compact objects, such   
as GRB's, XRF's possibly and blazars. XRF's show a clustering of   
the peak energy at $\sim 25 \keV$, which may be attributed to the same   
mechanism, provided the characteristic Lorentz factor, or the optical  
depth, is somewhat smaller than assumed here. Our results may also apply   
to blazars, where a clustering of the peak energies at $ 1-5 \MeV$   
is reported \citep{McNaron95}, if dissipation occurs at substantial   
optical depths and the bulk Lorentz factors are $\gtrsim 50$.  
    
\acknowledgements  
Research supported by NSF AST 0098416, 0307376, NASA NAFG5-13286.


\begin{thebibliography}{}  
  
\bibitem[Band \etal (1993)]{Band93}   
 Band, D., \etal 1993, \apj, 413, 281  
  
\bibitem[Brainerd (1998)]{Brainerd98}  
Brainerd, J.J. 1998, in the 19th Texas Symposium on Relativistic Astrophysics and  
Cosmology, Eds.: J. Paul, T. Montmerle, and E. Aubourg (Paris: CEA Saclay)  
  
\bibitem[Crider {\it et al.} (1997)]{Crider97}  
 Crider, A., \etal 1997, \apj, 479, L39  
  
\bibitem[Daigne \& Mochkovitch (1998)]{Daigne98}  
 Daigne, F. \& Mochkovitch, R. 1998, \mnras, 296, 275  
  
\bibitem[Frontera \etal (2000)]{Frontera00}  
 Frontera, F., \etal 2000, \apjs, 127, 59  
  
\bibitem [Ghirlanda \etal (2003)]{Ghirlanda03}  
 Ghirlanda, G., Celotti, A., \& Ghisellini, G. 2003, A\&A, 406, 879   
  
\bibitem [Ghisellini \& Celloti (1999)]{GC99}  
 Ghisellini, G., \& Celotti, A. 1999, \apj, 511, L93  
  
\bibitem [Ghisellini \etal (1988)]{GGS88}  
 Ghisellini, G., Guilbert, P.W., \& Svensson, R. 1988, \apj, 334, L5  
  
\bibitem[Guetta \etal (2004)]{Guetta04}  
 Guetta, D; Ghisellini, G; Lazzati, D; Celotti, A, 2004, A\&A, 421,  
 877  
  
\bibitem[Guilbert \etal (1983)]{GFR83}  
 Guilbert, P.W., Fabian, A.C., \& Rees, M.J. 1983, \mnras, 205, 593  
  
\bibitem[Lazzati \etal (2000)]{LGCR00}  
 Lazzati D, Ghisellini G, Celotti A \& Rees MJ 2000, \apj, 529, L17  
  
\bibitem[Lightman \& Zdziarski (1987)]{LZ87}  
 Lightman, A.P., \& Zdziarski, A.A. 1987, \apj, 319, 643  
  
\bibitem[McNaron-Brown \etal (1995)]{McNaron95}  
 McNaron-Brown, K., \etal 1995, \apj, 451, 575   
  
\bibitem[M\'esz\'aros (2002)]{Mes02}  
 M\'esz\'aros, P. 2002, ARA\&A, 40, 137  
  
\bibitem [M\'esz\'aros \etal (2002)]{MRRZ}  
 M\'esz\'aros, P, Ramirez-Ruiz, E, Rees, MJ, \& Zhang, B   
2002, \apj, 578, 812  
  
\bibitem [M\'esz\'aros \& Rees (2000)]{MR00}  
 M\'esz\'aros, P., \& Rees, M.J. 2000, \apj, 530, 292  
  
\bibitem[Pe'er \& Waxman (2004)]{PW04}  
 Pe'er, A., \& Waxman, E. 2004, \apj, 613, 448  
  
\bibitem[Pe'er \& Waxman (2004b) ]{PW05}  
 Pe'er, A., \& Waxman, E. 2004, \apj, in press (astro-ph/0409539)  
  
\bibitem[Pe'er \etal (2005)]{PM05}  
 Pe'er, A., M\'esz\'aros, P., \& Rees, M.J. 2005, in preparation  
  
\bibitem[Piran (2004)]{Piran04}  
 Piran, T. 2004, Rev. Mod. Phys. 76, 1143  
  
\bibitem[Preece \etal (1998)]{Preece98}  
 Preece, R., \etal 1998, \apj, 506, L23  
  
\bibitem[Preece \etal (2000)]{Preece00}  
 Preece, R., \etal 2000, \apj s.s., 126, 19  
  
\bibitem [Rees \& M\'esz\'aros (2004)]{MR04}  
 Rees, M.J. \& M\'esz\'aros, P., 2004, Ap.J. in press (astro-ph/0412702)  
  
\bibitem [Stern \& Poutanen (2004)]{SP04}  
 Stern, B.E., \& Poutanen, J. 2004, \mnras, 352, L35   
  
\bibitem [Svensson (1987)]{S87}  
 Svensson, R. 1987, \mnras, 227, 403  
  
\bibitem[Tavani (1996)]{Tavani96}  
 Tavani, M. 1996, \apj, 466, 768  
  
\bibitem[Waxman (2003)]{Waxman03}   
 Waxman, E.  2003, Gamma-Ray Bursts: The Underlying Model, in  
 Supernovae and Gamma-Ray bursters, Ed. K. Weiler    
 (Springer), Lecture Notes in Physics 598, 393--418.  
  
\end{thebibliography}
\end{document}